\documentclass[aps,prx,twocolumn,showpacs,amsmath,amssymb,floatfix,superscriptaddress,notitlepage]{revtex4-1}
\usepackage{graphicx}
\usepackage{amsmath}
\newcommand{\ket}[1]{|{#1}\rangle}
\newcommand{\bra}[1]{\langle{#1}|}
\newcommand{\braket}[1]{\langle{#1}}
\usepackage{amsfonts}

\usepackage[usenames,dvipsnames]{xcolor}
\usepackage[colorlinks=true,citecolor=Cerulean,linkcolor=RubineRed,urlcolor=Cerulean]{hyperref}

\begin{document}

\title{Shortcuts to Adiabaticity Assisted by Counterdiabatic Born-Oppenheimer Dynamics}

\author{Callum W. Duncan}
\affiliation{SUPA, Institute of Photonics and Quantum Sciences,
Heriot-Watt University, Edinburgh EH14 4AS, UK}
\affiliation{Department of Physics, University of Massachusetts, Boston, MA 02125, USA}
\author{Adolfo del Campo}
\affiliation{Department of Physics, University of Massachusetts, Boston, MA 02125, USA}
\affiliation{Theory Division, Los Alamos National Laboratory, MS-B213, Los Alamos, NM 87545, USA}
\begin{abstract}

Shortcuts to adiabaticity (STA) provide control protocols to guide  the dynamics of a quantum system through an adiabatic reference trajectory 
in an arbitrary prescheduled time. Designing STA proves challenging in complex quantum systems when the dynamics of the  degrees of freedom span different time scales. 
We introduce Counterdiabatic Born-Oppenheimer Dynamics  (CBOD) as a framework to design STA in systems with a large separation of energy scales.  CBOD exploits the Born-Oppenheimer approximation to separate the Hamiltonian into effective fast and slow degrees of freedom and calculate the corresponding counterdiabatic drivings for each subsystem. We show the validity of the CBOD technique via an example of coupled harmonic oscillators, which can be solved exactly for comparison, and further apply it to a system of two-charged particles. 
\end{abstract}

\maketitle

\section{Introduction}

Tailoring the nonadiabatic dynamics of quantum matter is an open problem at the frontiers of physics with important applications in emergent quantum technologies.
Control protocols relying on adiabatic dynamics are natural to prescribe the evolution of a system along a reference adiabatic trajectory. While they are robust against uncontrolled errors in the experimental implementation, they are susceptible to decoherence. Driving protocols known as shortcuts to adiabaticity (STA) provide an alternative, by  speeding up an adiabatic reference trajectory of a quantum system in a prescheduled amount of time \cite{Chen2010a}.

STA have found broad applications in quantum mechanical systems of varying complexity. They can be used to guide the dynamics of systems  with a discrete energy spectrum \cite{Demirplak2003,Demirplak2005,Berry2009,Chen2010b,Ruschhaupt2012,Masuda2015}, as shown in the laboratory \cite{Bason2012,Zhang13,Du2016,Zhou2017}.
Similarly, STA can be used to control the degrees of freedom of continuous variables systems \cite{Masuda2009,Chen2010a,Torrontegui2011,Choi2011,Choi2011b,Choi2013,Deffner2014,Patra2017,Jarzynski2017} as demonstrated by the fast control of a trapped ion in phase space \cite{An2016}.
In the context of trapped ultracold atoms, early theoretical results indicated that STA could be applied to many-body systems \cite{Campo2011,Campo2012b,Campo2013,Deffner2014}.  Ultrafast expansions and compressions of  atomic clouds have by now been implemented in a wide variety of interaction regimes including thermal clouds \cite{Schaff2010}, Bose-Einstein condensates well described by mean field theory \cite{Schaff2011a,Schaff2011}, tightly confined quasi-one-dimensional atomic clouds with phase fluctuations \cite{Rohringer2015},  and a unitary Fermi gas as a paradigmatic instance of a strongly-coupled quantum fluid \cite{Deng2016,Deng2018}.
In addition, theoretical studies have shown that  STA can be used to guide the evolution of many-body quantum systems  that exhibit quantum critical behavior \cite{Campo2012a,Campo2014b}. In this context, STA can be used to  suppress  excitation formation across a phase transition \cite{Campo2015}. Implementing STA may require modifying the systems Hamiltonian with nonlocal interactions including high order terms \cite{Campo2012a,Damski2014}. However, 
the required controls may be simplified or absorbed into the form of the original system Hamiltonian \cite{Takahashi2013a,Saberi2014,Mukherjee2016,Takahashi2017}.
Further efforts to control the dynamics of many-body systems  have been put forward relying on integrability (e.g. the existence of Lax pairs) \cite{Okuyama2016} or variational methods \cite{Sels2017}.
Despite this surge of progress, applications of STA remain mostly confined to systems with few degrees of freedom or the control of certain collective modes in many-body quantum systems. 

Designing STA requires the ability to control and describe the time-evolution of a system, an  ubiquitous  challenge across a variety of fields when dealing with complex quantum systems. 
Among them, a prominent instance occurs in quantum chemistry, in the study of  quantum systems with degrees of freedom spanning different  time and energy scales \cite{Kouppel2007}. When the separation of scales is sufficiently large, it is possible to decouple the dynamics via the Born-Oppenheimer approximation (BOA) \cite{born1924}. Born and Oppenheimer considered the description of a molecule and harnessed the separation of energy scales  between the  electronic and nuclear rotational and vibrational motions to simplify the description.
As the electronic mass is much smaller than that of the atomic nuclei, the motion of the corresponding degrees of freedom occurs  on vastly different time scales. 
When the mass ratio is large enough, the electrons move in an effectively static configuration of the nuclei. Further, this separation leads to the evolution of the nuclear component  in the presence of a potential set by the energy of the electronic motion. When the assumptions of the approximation are relaxed, the nuclear motion is subject to a Berry vector potential due to the electronic motion \cite{Moody1986,Berry1990,Bohm1992,Mead1992,min2014}.
Understanding the BOA and its limits has proved particularly fruitful in the field of spectroscopy \cite{piela2013,Parson2015}, the study of molecular dynamics, and in  computational models \cite{Payne1992,Barnett1993,Butler1998,Kuo2006,Niklasson2006,Pisana2007,Niklasson2008,Odell2011,Cawkwell2012,Lin2013}. There has also been recent advances in the quantum simulation of molecular dynamics, with the recent proposal and experimental implementation of vibrational spectroscopy with trapped ions via boson sampling \cite{Huh2015,Huh2017,Shen2017}.

In this paper, we introduce Counterdiabatic Born-Oppenheimer Dynamics  (CBOD) as an efficient  technique for the fast control of complex systems that are well described by the BOA. To this end, we first provide a brief  summary of the BOA and the engineering of STA by counterdiabatic driving in sections \ref{BOMsec} and \ref{CDsec}, respectively. We then present CBOD in \ref{CBODsec} and demonstrate its validity with a paradigmatic example of coupled harmonic oscillators of unequal mass, in \ref{CBODex_sec}. We end by considering an example of a trapped particle and free particle interacting via a Coulomb-like term in \ref{CBODexCharge_sec}.


\section{Born-Oppenheimer method}\label{BOMsec}
In order to fix the notation, we will briefly discuss the Born-Oppenheimer approximation (BOA) that will be at the core of the CBOD technique. Throughout this work, we will consider two-body systems. However, the approach can be readily generalized to $N$-body systems with two sub-sets of slow (heavy) and fast (light) particles, that we  indicate with the labels $S$ and $F$, respectively. 
We will make the assumption that $m_S  \gg m_F$. 
In this section, we will introduce both the conventional BOA as well as the relaxed BOA, whereby, the fast degrees of freedom give rise to a Berry vector potential for the slow variables.

\subsection{Conventional Born-Oppenheimer approximation}
\label{sec:BOA}

Consider a  Hamiltonian of the form
\begin{align}
\hat{H} = \frac{\hat{\mathbf{p}}_S^2}{2 m_S} + \frac{\hat{\mathbf{p}}_F^2}{2 m_F} + V\left(\hat{\mathbf{x}}_S,\hat{\mathbf{x}}_F\right),
\end{align}
where $\hat{\mathbf{p}}_i = - i \hbar \nabla_i$ ($i=S,F$) denotes the momentum operator and $V\left(\hat{\mathbf{x}}_S,\hat{\mathbf{x}}_F\right)$ is a global  potential term. The latter can generally be decomposed as  
\begin{align}
V\left(\hat{\mathbf{x}}_S,\hat{\mathbf{x}}_F\right) = V_{S}\left(\hat{\mathbf{x}}_S\right) + V_{F}\left(\hat{\mathbf{x}}_F\right) + V_{I}\left(\hat{\mathbf{x}}_S,\hat{\mathbf{x}}_F\right),
\end{align}
with the first two terms acting exclusively on the slow and fast coordinates, respectively, and an interaction term, which is not separable in the coordinate  representation in terms of $\{x_S,x_F\}$.  
The general form given by this Hamiltonian includes the usual molecular Hamiltonian for electronic and nuclear dynamics, frequently used in quantum chemistry \cite{born1924,Schwabl2007,piela2013}.

We wish to obtain the solutions to the Schr\"{o}dinger equation
\begin{align}
\left[\frac{\hat{\mathbf{p}}_S^2}{2 m_S} + \frac{\hat{\mathbf{p}}_F^2}{2 m_F} + V\left(\hat{\mathbf{x}}_S,\hat{\mathbf{x}}_F\right)\right]\Psi = E \, \Psi.\label{eq:FullSE}
\end{align}
Given the difference in mass ($m_S\gg m_F$), the  BOA suppresses the kinetic energy term of the slow variables \cite{weinberg2015Book} to obtain a reduced Hamiltonian for the fast ones. As $\hat{\mathbf{x}}_S$ commutes with this reduced Hamiltonian, we can simultaneously obtain the solutions of the reduced and full sub-systems. The reduced Hamiltonian governs the Schr\"{o}dinger equation  of the  fast sub-system
\begin{align}
\left[ \frac{\hat{\mathbf{p}}_F^2}{2 m_F} + V\left(\mathbf{x}_S,\hat{\mathbf{x}}_F\right)\right]\phi_n\left(\mathbf{x}_F;\mathbf{x}_S\right) = \varepsilon_n\left(\mathbf{x}_S\right) \phi_n\left(\mathbf{x}_F;\mathbf{x}_S\right).\label{eq:ReducedSE}
\end{align}
The slow coordinates $\mathbf{x}_S$ can be regarded as a parameter on which the reduced system eigenvalues and eigenvectors depend. The solutions to the reduced Schr\"{o}dinger equation form a complete set, in terms of which the full solution to the complete Schr\"{o}dinger equation, Eq.~\eqref{eq:FullSE}, can be written as
\begin{align}
\Psi = \sum_n \phi_n\left(\mathbf{x}_F;\mathbf{x}_S\right) \psi_n\left(\mathbf{x}_S\right),\label{eq:States}
\end{align}
where $n$ runs over the eigenstates of the reduced Hamiltonian. We will assume in this work that the fast sub-system is in a single eigenstate $n$, avoiding the need for the summation in Eq. (\ref{eq:States}).

Using the product expansion of the wave function, Eq.~\eqref{eq:States}, and the reduced Schr\"{o}dinger equation, Eq.~\eqref{eq:ReducedSE}, the full Schr\"{o}dinger equation reads
\begin{align}
\left[\frac{\hat{\mathbf{p}}_S^2}{2 m_S} + \varepsilon_n\left(\hat{\mathbf{x}}_S\right)\right] & \phi\left(\mathbf{x}_F;\mathbf{x}_S\right) \psi\left(\mathbf{x}_S\right) \nonumber \\ & = E \phi\left(\mathbf{x}_F;\mathbf{x}_S\right) \psi\left(\mathbf{x}_S\right).\label{eq:FullBothStates}
\end{align}
In the conventional BOA the derivatives of the fast sub-system wave function, $\phi\left(\mathbf{x}_F;\mathbf{x}_S\right)$, with respect to $\mathbf{x}_S$ are neglected in the above equation, i.e.,
\begin{align}
\frac{\hat{\mathbf{p}}_S^2}{2 m_S} \phi\left(\mathbf{x}_F;\mathbf{x}_S\right) \psi\left(\mathbf{x}_S\right) \approx \phi\left(\mathbf{x}_F;\mathbf{x}_S\right) \frac{\hat{\mathbf{p}}_S^2}{2 m_S} \psi\left(\mathbf{x}_S\right).
\end{align}
Integrating out the fast degrees of freedom is then straightforward and leads to a slow sub-system Schr\"{o}dinger equation in the final form
\begin{align}
\left[\frac{\hat{\mathbf{p}}_S^2}{2 m_S} + \varepsilon_n\left(\hat{\mathbf{x}}_S\right)\right]  \psi\left(\mathbf{x}_S\right) = E\, \psi\left(\mathbf{x}_S\right).
\end{align}
Note, $E$ gives the full energy of the system while the full approximate wave function is given by Eq.~\eqref{eq:States}.

The conventional BOA  involves truly two approximations: (1) the energy scales of the system are vastly different allowing for the suppression of one of the kinetic energies to obtain the reduced Hamiltonian and (2) corrections due to the elimination of derivatives in $\mathbf{x}_S$ of the reduced wave function are small. These corrections are referred to as {\it diagonal corrections} and they are usually negligible in comparison to the energy scale of the fast sub-system \cite{piela2013}. The form of these corrections and their calculation is a vibrant area of research in its own right \cite{Handy1986,Fernandez1994,Handy1996,Csaszar1998,Valeev2003,Gauss2006,Helgaker2008,Gherib2016}.
We have added another approximation to the conventional approach, that the fast sub-system evolves adiabatically, this was invoked when the sum over the fast sub-system states was neglected. This is a common approximation when using the BOA to describe time-evolution  \cite{piela2013,Hagedorn1980}, as the fast sub-system is assumed to quickly relax to its ground state in the time-scale of the slow motion.  Note, that when we combine the BOA with counterdiabatic driving, we will assume the fast sub-system either evolves adiabatically or, more importantly for our approach, that the fast sub-system is driven such that adiabaticity is enforced; in either case, the adiabatic approximation is met. 

\subsection{Relaxed Born-Oppenheimer Approximation}
\label{sec:GeneralBO}

The approximation involving the elimination of the derivatives with respect to $\mathbf{x}_S$ made in the conventional BOA can be relaxed \cite{Mead1992,Bohm1992}. Generally, the neglected diagonal term can couple arbitrary eigenfunctions of the fast degrees of freedom. A relaxed BOA consists of keeping the resulting cross-terms of the $\mathbf{x}_S$ derivative while neglecting transitions between these different eigenstates. This leads to  the appearance of a Berry connection between the two sub-systems in the full Schr\"{o}dinger equation, which plays a role analogous to the vector potential in the quantum mechanics of a charged particle in an electromagnetic field \cite{Mead1992,Bohm1992,Wilczek1989}. 

Starting from the Schr\"{o}dinger equation \eqref{eq:FullBothStates}, our goal is to obtain a Schr\"{o}dinger equation which is solely dependent on the slow degree of freedom, without invoking the previous approximation that disregards the derivatives of the momentum operator. To remove the fast degree of freedom we multiply  Eq.  \eqref{eq:FullBothStates} from the left by $\phi\left(\mathbf{x}_F;\mathbf{x}_S\right)^\dagger$ and integrate over $\mathbf{x}_F$  to obtain 
\begin{align}
\int d\mathbf{x}_F \phi\left(\mathbf{x}_F;\mathbf{x}_S\right)^\dagger \left[ \frac{\hat{\mathbf{p}}_S^2}{2 m_S} + \varepsilon_n\left(\hat{\mathbf{x}}_S\right) \right] \phi\left(\mathbf{x}_F;\mathbf{x}_S\right) \psi\left(\mathbf{x}_S\right) & \nonumber\\ = \int d\mathbf{x}_F \phi\left(\mathbf{x}_F;\mathbf{x}_S\right)^\dagger E \phi\left(\mathbf{x}_F;\mathbf{x}_S\right) \psi\left(\mathbf{x}_S\right).
\end{align}
There is only one non-trivial integral in the above Schr\"{o}dinger equation, which is,
\begin{align}
\int d\mathbf{x}_F \phi\left(\mathbf{x}_F;\mathbf{x}_S\right)^\dagger \frac{\hat{\mathbf{p}}_S^2}{2 m_S} \phi\left(\mathbf{x}_F;\mathbf{x}_S\right) \psi\left(\mathbf{x}_S\right).
\end{align}
It is an algebraic exercise \cite{weinberg2015Book} to obtain the terms arising from  this integral, which can be written in terms of a vector and a scalar potential in the slow sub-system Schr\"{o}dinger equation 
\begin{align}
\left[ -\frac{\hbar^2}{2m_S} \left(\nabla_{\mathbf{x}_S} - i \mathcal{A}\left(\mathbf{x}_S\right) \right)^2 +\frac{\hbar^2}{2m_S} g(\mathbf{x}_S)  + \varepsilon_n\left(\hat{\mathbf{x}}_S\right) \right] & \psi\left(\mathbf{x}_S\right) \nonumber \\  = E\,  \psi\left(\mathbf{x}_S\right)& \label{eq:BOSchrodingerVector}
\end{align}
with
\begin{align}
\mathcal{A}\left(\mathbf{x}_S\right) & = i \int d\mathbf{x}_F \phi\left(\mathbf{x}_F;\mathbf{x}_S\right)^\dagger \nabla_{\mathbf{x}_S} \phi\left(\mathbf{x}_F;\mathbf{x}_S\right) \label{eq:GeneralBOVector}\nonumber\\
& = i \langle \phi|\nabla_{\mathbf{x}_S}\phi\rangle,\\
g(\mathbf{x}_S) & = \int d\mathbf{x}_F \left[\nabla_{\mathbf{x}_S} \phi\left(\mathbf{x}_F;\mathbf{x}_S\right)^\dagger\right] \nabla_{\mathbf{x}_S} \phi\left(\mathbf{x}_F;\mathbf{x}_S\right) \nonumber \\ & + \left[\int d\mathbf{x}_F \phi\left(\mathbf{x}_F;\mathbf{x}_S\right)^\dagger \nabla_{\mathbf{x}_S} \phi\left(\mathbf{x}_F;\mathbf{x}_S\right)\right]^2\nonumber\\
&=\langle \nabla_{\mathbf{x}_S}\phi|\nabla_{\mathbf{x}_S}\phi\rangle-\langle \nabla_{\mathbf{x}_S}\phi|\phi\rangle\langle \phi|\nabla_{\mathbf{x}_S}\phi\rangle. \label{eq:GeneralBOScalar}
\end{align}
The vector potential $\mathcal{A}\left(\mathbf{x}_S\right)$ is the familiar Berry connection. The  scalar potential $\varepsilon_n\left(\hat{\mathbf{x}}_S\right)$ is local and is  dictated by the fast variables. In addition, there is a contribution to the  scalar potential experienced by the slow degrees of freedom  that is given by $g(\mathbf{x}_S)$, which is 
the trace of the quantum geometric tensor  \cite{Provost1980} associated with the change of the eigenstates $|\phi\rangle$ with respect to the  slow coordinates $\mathbf{x}_S$, treated as a parameter.


\section{Counter-diabatic driving}\label{CDsec}

Among the variety of techniques available to engineer STA, counterdiabatic driving (CD) stands out as a universal approach. It relies on the use of  auxiliary counter-diabatic fields to  guide the evolution of the quantum system of interest through an adiabatic reference trajectory. CD was  developed in the context of molecular dynamics by Demirplak and Rice \cite{Demirplak2003,Demirplak2005,Demirplak2008}, as an alternative to strictly adiabatic population transfers between molecular states; see also the independent and closely related work by Berry \cite{Berry2009}.  CD and related protocols have been recently  implemented in a variety of platforms for quantum technologies including  trapped ions \cite{An2016}, nitrogen-vacancy centres in diamond \cite{Zhang13,Zhou2017}, ultracold atoms in optical lattices \cite{Bason2012} and as a method to speed-up stimulated Raman adiabatic passage in ultracold gases \cite{Du2016}.
It has become a popular technique to control and engineer the nonadiabatic evolution of quantum systems while enforcing the following of adiabatic trajectories \cite{Demirplak2003,Demirplak2005,Demirplak2008,Berry2009,Campo2013,Sels2017}.

Counterdiabatic driving relies on the spectral properties, eigenstates and energies, of the driven Hamiltonian of interest
\begin{align}
\hat{H}_0 \left(t\right) \ket{n\left(t\right)} = \varepsilon_n \left(t\right) \ket{n\left(t\right)}.
\end{align}
According to the adiabatic approximation, the state of a system prepared in an eigenstate $\ket{n\left(0\right)}$ at $t=0$  evolves under  a slowly-varying $\hat{H}_0 \left(t\right)$ into
\begin{eqnarray}
\ket{\psi_n^{\rm ad} \left(t\right)} &=& \exp\bigg[ - \frac{i}{\hbar} \int_0^t dt^\prime \varepsilon_n \left(t^\prime\right) \nonumber \\ & &- \int_0^t dt^\prime \braket{n\left(t^\prime\right)} \ket{\partial_{t^\prime} n\left(t^\prime\right)} \bigg] \ket{n\left(t\right)}, 
\label{adiabpsi}
\end{eqnarray}
which includes the dynamical phase as well as the geometric phase associated with the Berry connection $i\braket{n\left(t^\prime\right) \ket{\partial_{t^\prime} n\left(t^\prime\right)}}$.

A STA protocol assisted by CD can be designed by identifying  a modified driven Hamiltonian $\hat{H}\left(t\right)$ such that the adiabatic evolution  (\ref{adiabpsi})  becomes the exact solution of the corresponding time-dependent Schr\"{o}dinger equation 
\begin{eqnarray}
\hat{H}\left(t\right) \ket{\psi_n^{\rm ad} \left(t\right)}  = i \hbar \partial_t \ket{\psi_n^{\rm ad} \left(t\right)} .
\label{tdsecd}
\end{eqnarray}
Hence, no matter how fast the system is driven, the evolution is described by  the adiabatic trajectory  (\ref{adiabpsi}), i.e., without the requirement of slow driving.
The corresponding time-evolution operator
also fulfills (\ref{tdsecd}),
which allows the identification of the  modified driven Hamiltonian as the generator of evolution
\begin{eqnarray}
\hat{H}\left(t\right) = i \hbar \left[\partial_t \hat{U} \left(t\right)\right] \hat{U}^\dagger \left(t\right).
\end{eqnarray}
Making use of the following form of the  time-evolution operator 
\begin{align}
\hat{U}\left(t\right) = & \sum_n \ket{\psi_n^{\rm ad} \left(t\right)}  \bra{n\left(0\right)},
\end{align}
the modified driven Hamiltonian is found by explicit computation  \cite{Demirplak2003,Berry2009}
\begin{eqnarray}
\hat{H}\left(t\right) = \sum_n \varepsilon_n \ket{n} \bra{n} +\hat{H}_1 \left(t\right),
\label{Hfull}
\end{eqnarray}
where we have defined
\begin{eqnarray}
\label{eq:CDHamiltonian1}
\hat{H}_1 \left(t\right) = i \hbar \sum_n \left( \ket{\partial_t n} \bra{n} - \braket{n}\ket{\partial_t n} \ket{n} \bra{n} \right) .
\end{eqnarray}
The first term in (\ref{Hfull}) is recognized as the spectral decomposition of the original system Hamiltonian $\hat{H}_0 \left(t\right)$. 
The second term,  $ \hat{H}_1$,  is  the auxiliary  CD  term required so that  the adiabatic trajectory $\ket{\psi_n^{\rm ad} \left(t\right)}$ in Eq. (\ref{adiabpsi}) becomes an exact solution of (\ref{tdsecd}), that is the Schr\"odinger equation for the full driving hamiltonian $\hat{H}=\hat{H}_0+\hat{H}_1$.

When the  energy spectrum of $\hat{H}_0$ is  non-degenerate, the additional CD term can be recasted using the differential of the time-independent Schr\"{o}dinger equation of the original system Hamiltonian \cite{Berry2009}, $\hat{H}_0\left(t\right)$,
\begin{align}
\braket{m\left(t\right)} \ket{\partial_t n\left(t\right)} = \frac{\bra{m\left(t\right)} \partial_t \hat{H}_0\left(t\right) \ket{n\left(t\right)}}{\varepsilon_n\left(t\right) - \varepsilon_m\left(t\right)},
\end{align}
which  yields the following alternative expression for the auxiliary CD term
\begin{align}
\hat{H}_1\left(t\right) = i \hbar \sum_{m \neq n} \sum_n \frac{\hat{P}_m(t) \partial_t \hat{H}_0\left(t\right) \hat{P}_n(t)}{\varepsilon_n\left(t\right) - \varepsilon_m\left(t\right)},\label{eq:CDHamiltonian2}
\end{align}
in terms of the projector $\hat{P}_m(t)=\ket{m\left(t\right)}\bra{m\left(t\right)}$.


\section{Counterdiabatic Born-Oppenheimer Dynamics  (CBOD)}\label{CBODsec}

The Born-Oppenheimer method and the theory of counterdiabatic driving can be exploited jointly to engineer the fast nonadiabatic control of complex systems, as we next discuss. The counterdiabatic drivings for the slow and fast sub-systems can be obtained via the BOA, either via the conventional or relaxed variants. These auxiliary control terms can then be used to drive  the (exact) system  Hamiltonian, a technique we shall term as Counterdiabatic Born-Oppenheimer Dynamics (CBOD).

\subsection{CD with the conventional Born-Oppenheimer approximation}

We first consider the conventional BOA, according to which the  fast and slow sub-system Hamiltonians are 
\begin{eqnarray}
\hat{H}_F\left(\hat{\mathbf{x}}_F,t;\mathbf{x}_S\right) &=& \frac{\hat{\mathbf{p}}_F^2}{2 m_F} + V\left(\mathbf{x}_S,\hat{\mathbf{x}}_F\right),\\
\hat{H}_{S}\left(\hat{\mathbf{x}}_S,t\right) &= &\frac{\hat{\mathbf{p}}_S^2}{2 m_S} + \varepsilon_n\left(\hat{\mathbf{x}}_S\right). 
\end{eqnarray}
The required CD terms can be found via the general expression Eq.~\eqref{eq:CDHamiltonian1} and  for the slow and fast sub-systems read respectively
\begin{eqnarray}
\hat{H}_{F,1} \left(t\right) &=& i \hbar \bigg( \ket{\partial_t \phi} \bra{\phi} - \braket{\phi}\ket{\partial_t \phi} \ket{\phi} \bra{\phi} \bigg), \label{eq:CBODFast}\\
\hat{H}_{S,1} \left(t\right) &=& i \hbar \bigg( \ket{\partial_t \psi} \bra{\psi} - \braket{\psi}\ket{\partial_t \psi} \ket{\psi} \bra{\psi} \bigg).\label{eq:CBODSlow}
\end{eqnarray}
By contrast,  the CD term for the full system is
\begin{align}
\hat{H}_{\mathrm{Full},1} \left(t\right) = i \hbar \bigg( \ket{\partial_t \Psi} \bra{\Psi} - \braket{\Psi}\ket{\partial_t \Psi} \ket{\Psi} \bra{\Psi} \bigg).\label{eq:CDFull1}
\end{align}
The full form can be rewritten, exploiting the tensor product structure of the full wave function $\ket{\Psi} = \ket{\phi} \otimes \ket{\psi}$ (i.e. separable in the BOA). We note that while this separable form is natural in the BOA, it can be invoked generally \cite{Hunter75,Abedi10}.
Substituting in the factored form of the full wave function, the full CD can be written as
\begin{eqnarray}
\hat{H}_{\mathrm{Full},1} \left(t\right) & = &i \hbar \bigg( \ket{\partial_t \phi} \bra{\phi}\otimes \ket{\psi} \bra{\psi} + \ket{\phi} \bra{\phi} \otimes\ket{\partial_t \psi} \bra{\psi} 
\nonumber\\ & & - \braket{\phi}\ket{\partial_t \phi}\otimes \ket{\psi} \bra{\psi} - \ket{\phi} \bra{\phi}\otimes \braket{\psi}\ket{\partial_t \psi}\bigg) 
\nonumber\\ & \equiv & \hat{H}_{F,1} \left(t\right)\otimes \ket{\psi}\bra{\psi} + \ket{\phi}\bra{\phi} \otimes\hat{H}_{S,1}\left(t\right).
\end{eqnarray}
Therefore, the CBOD technique  simplifies the global CD control by driving  the two sub-systems separately, this is, by using the auxiliary control terms (\ref{eq:CBODFast}) and (\ref{eq:CBODSlow}) as opposed to (\ref{eq:CDFull1}).

We can gain further insight into the CBOD terms by assuming the spectra of $\hat{H}_{S}$ and $\hat{H}_{F}$ to be nondegenerate,  as this allows the recasting of the CD control terms into the form of Eq.~\eqref{eq:CDHamiltonian2},
\begin{align}
\hat{H}_{F,1} & = i \hbar \sum_{m \neq n} \sum_n \frac{\ket{\phi_m}\bra{\phi_m} \partial_t V\left(\mathbf{x}_S,\hat{\mathbf{x}}_F\right)\ket{\phi_n}\bra{\phi_n}}{\varepsilon_n\left(\mathbf{x}_S\right) - \varepsilon_m\left(\mathbf{x}_S\right)},\label{eq:FastCD}\\
\hat{H}_{S,1} & = i \hbar \sum_{m \neq n} \sum_n \frac{\ket{\psi_m}\bra{\psi_m} \partial_t \varepsilon_n\left(\hat{\mathbf{x}}_S\right) \ket{\psi_n}\bra{\psi_n}}{E_n - E_m}. \label{eq:SlowCD}
\end{align}
Note, that in the fast control $\hat{H}_{F,1}$, $\mathbf{x}_S$ is treated as a parameter, which is the case of the reduced Hamiltonian in Eq.~\eqref{eq:ReducedSE} in the BOA. 

As customary  in STA  assisted by CD,   the required auxiliary terms are off-diagonal  in state space. However, it is useful to focus on 
the dependencies on $\mathbf{x}_S$ and $\mathbf{x}_F$. We wish to compare the CBOD terms with the exact CD term without resorting to the Born-Oppenheimer approximation. Assuming the system is exactly solvable the CD is
\begin{align}
\hat{H}_{\mathrm{exact},1} = i \hbar \sum_{m \neq n} \sum_n \frac{\ket{\chi_m}\bra{\chi_m} \partial_t V\left(\hat{\mathbf{x}}_S,\hat{\mathbf{x}}_F\right) \ket{\chi_n}\bra{\chi_n}}{\epsilon_n - \epsilon_m},\label{eq:ExactCD}
\end{align}
with $\hat{H}_0 \ket{\chi_n} = \epsilon_n \ket{\chi_n}$, $\ket{\chi}$ the exact eigenstates and $\epsilon_n$ the exact eigenvalues. Due to the interactions between slow and fast degrees of freedom, which are not separable in the $\{\mathbf{x}_S,\mathbf{x}_F\}$ space, the operator $\hat{H}_{\mathrm{exact},1} $ can be complicated to implement, as it involves a generally complex coupling between the two sub-spaces (or two particles). 

Therefore, CBOD provides a computational advantage over the exact STA approach. Indeed, it circumvents the need to deal with the full spectra,  proceeding, instead, in two-subsequent steps; treating first the fast degrees of freedom and then the slow ones.  CBOD thus benefits from the dimensional reduction of the problem to engineer the CD term for the  fast sub-system driving,  Eq.~\eqref{eq:FastCD}.

In addition, CBOD may simplify the required drivings by potentially removing or reducing the coupling between the two sub-spaces. In such a case,  CBOD controls will be simpler to implement than the exact CD terms. For the slow sub-system, Eq.~\eqref{eq:SlowCD},  the off-diagonal terms are coupled via the time derivative of the energy for the fast sub-system. This is the potential surface which the slow sub-system  experiences due to the fast sub-system, and it is of no surprise that to enforce adiabatic evolution it is required to drive off-diagonal terms with this as the scaling. The coordinate dependence of slow sub-system counterdiabatic driving, Eq.~\eqref{eq:SlowCD} is simplified in comparison to the exact driving, Eq.~\eqref{eq:ExactCD}, as the potential energy surfaces can only depend on $\mathbf{x}_S$. Therefore, the slow sub-system driving only requires operators which act on the slow sub-system space and will have no cross terms between the two sub-spaces. Therefore, CBOD readily simplifies the required control term for the slow sub-system, without further approximations.

\subsection{CD with the relaxed Born-Oppenheimer approximation}

In what follows we derive the modified driving controls when the relaxed BOA is used, as discussed in Sec.~\ref{sec:GeneralBO}. The required CD terms for the fast and exact systems are the same as in Eqs.~\eqref{eq:CBODFast} and~\eqref{eq:CDFull1}; alternatively, by \eqref{eq:FastCD} and \eqref{eq:ExactCD}. The slow sub-system has a modified Hamiltonian resembling  that of a particle in an electromagnetic field, see Eq.~\eqref{eq:BOSchrodingerVector}.  

To obtain the CD under the assumption of no degeneracies in the spectra according to Eq.~\eqref{eq:CDHamiltonian2}, we first need to obtain the time derivative of the Hamiltonian 
\begin{align}
\partial_t \hat{H}_S = & \frac{\hbar^2}{2m_S} \Big[2i\left(\partial_t \mathcal{A}\right) \nabla_{\mathbf{x}_S} + 2 \mathcal{A} \left(\partial_t \mathcal{A}\right) \nonumber +  i\left( \nabla_{\mathbf{x}_S} \partial_t \mathcal{A}\right) \Big] \\ &  +  \frac{\hbar^2}{2m_S} \partial_t g_n + \partial_t \varepsilon_n.
\end{align}
The latter admits the compact form
\begin{align}
\partial_t & \hat{H}_S\left(\mathbf{x}_S,t\right) = \frac{i \hbar}{m_S} \{ \dot{\mathcal{A}}_n\left(\hat{\mathbf{x}}_S,t\right), \nabla_{\mathbf{x}_S} \} + \mathcal{V}_n \left(\hat{\mathbf{x}}_S,t\right),
\end{align}
with the potential term given by
\begin{eqnarray}
\mathcal{V}_n \left(\hat{\mathbf{x}}_S,t\right) & = &\frac{\hbar^2}{m_S} \mathcal{A}_n\left(\hat{\mathbf{x}}_S,t\right) \dot{\mathcal{A}}_n\left(\hat{\mathbf{x}}_S,t\right) \nonumber\\ 
& & - \frac{i \hbar^2}{2m_S}  \nabla_{\hat{\mathbf{x}}_S} \dot{ \mathcal{A}}_n\left(\hat{\mathbf{x}}_S,t\right) 
\nonumber\\
& & +\frac{\hbar^2}{2m_S}  \partial_t g_n\left(\hat{\mathbf{x}}_S,t\right)+ \partial_t \varepsilon_n\left(\hat{\mathbf{x}}_S,t\right),
\end{eqnarray}
where $\{ \cdot , \cdot \}$ denotes the anti-commutator and $\dot{\mathcal{A}} \equiv \partial_t \mathcal{A}$. The slow sub-system counterdiabatic driving is therefore given by
\begin{align}
&\hat{H}_{S,1}  = i \hbar \sum_{m \neq n} \sum_n \frac{1}{E_n - E_m} \\ & \times \Big[ \hat{P}_{m,S}\bigg( \frac{i \hbar}{m_S} \{ \dot{\mathcal{A}}_n\left(\hat{\mathbf{x}}_S,t\right), \nabla_{\mathbf{x}_S} \} + \mathcal{V}_n \left(\hat{\mathbf{x}}_S,t\right) \bigg)  \hat{P}_{n,S} \big],\nonumber
\end{align}
where $ \hat{P}_{m,S}= \ket{m_S}\bra{m_S}$.
As a result, even with the relaxed BOA, the slow sub-system CD term only depends on operators related to the slow coordinate $\mathbf{x}_S$. 

\subsection{Applicability of Counterdiabatic Born-Oppenheimer Dynamics (CBOD)}

CBOD, as an approximate technique, does not necessarily enforce the evolution of the system Hamiltonian to follow the adiabatic trajectory exactly. It resorts to the counterdiabatic driving terms constructed via the BOA to drive the (exact) system, which includes couplings between slow and fast subsystems beyond BOA. Said differently, CBOD is constructed to drive the  fast and slow Hamiltonians of the BO Hamiltonian (as opposed to the exact system Hamiltonian) exactly through the adiabatic manifold. 
 We will assess the validity of CBOD using the fidelity between the resulting state and  that of the exact adiabatic evolution after a modulation of the system Hamiltonian in a prescheduled time. 

The implementation of the CBOD technique, in general, involves the  following steps:
\begin{enumerate}
\item Check the validity of BOA, i.e., that there are two separated energy scales in the region of interest.

\item Derive the counterdiabatic drivings using the BOA.

\item Apply these (approximate) control terms  to guide the dynamics of the (exact) system Hamiltonian.
\end{enumerate}
We will consider next an example discussing each step in detail and certify CBOD by comparing its performance to the exact CD evolution.


\section{Coupled harmonic system}\label{CBODex_sec}

To illustrate CBOD, we next consider the engineering of STA to drive two coupled harmonic oscillators, that can represent, e.g.,  two atoms in a harmonic trap interacting via a spring-like term. This model  has been previously used to assess the BOA \cite{Fernandez94},  admits an exact solution \cite{Estes1968,Han1999,Plenio2004,Kao2016,Makarov2017,Brandt2012} and is realizable in controllable quantum systems of ion traps \cite{Brown2011,Harlander2011,Blatt2012,Schneider2012}. It, therefore, constitutes a natural test-bed for CBOD.

\begin{figure*}[t]
\begin{center}
\includegraphics[width=0.95\linewidth]{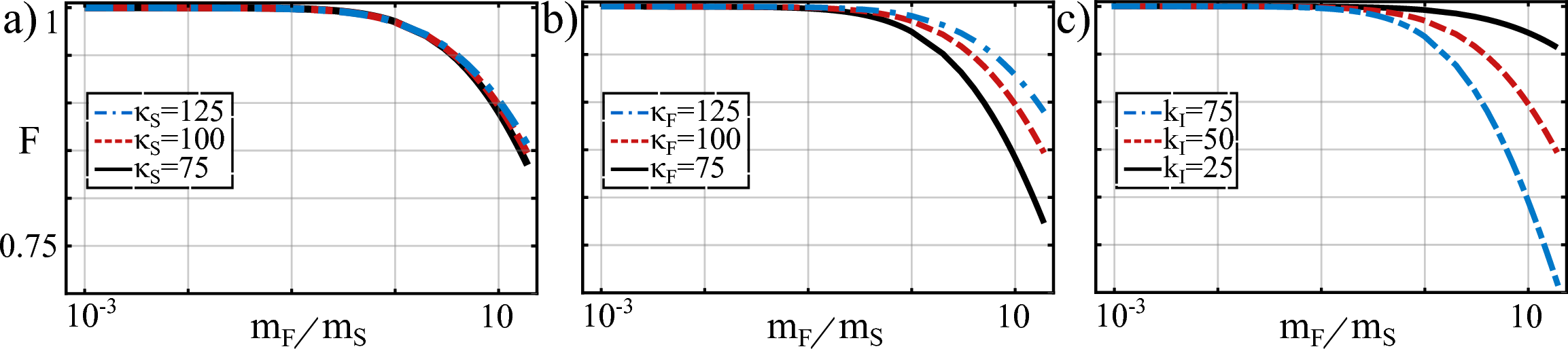}
\end{center}
\caption{\label{fig:Fidelityk} 
{\bf Ground state fidelity between the exact and BOA wave functions.}
Fidelity between the exact ground state wave function of two-coupled harmonic oscillators and the corresponding BOA wave function, as a function of the  mass ratio in a log-scale. a)  Increasing the slow spring constant, enhances the accuracy of BOA ($\kappa_F = 100$, $k_I=50$). b) Similarly, increasing the fast spring constant increases the fidelity ($\kappa_S = 100$, $k_I=50$). c) As the the coupling between the slow and fast subsystems is enhanced by increasing the interaction spring constant, the BOA begins to break down (with $\kappa_S = 100$ and $\kappa_F=100$).}
\end{figure*}

Specifically,  we consider the Hamiltonian 
\begin{eqnarray}
\hat{H}_0\left(t\right)&  = & \frac{\hat{p}^2_S}{2 m_S} + \frac{\hat{p}^2_F}{2 m_F} + \frac{1}{2} k_S\left(t\right)\hat{x}_S^2 + \frac{1}{2} k_F\left(t\right)\hat{x}_F^2 \nonumber 
\\ & & + \frac{1}{2} k_I\left(t\right)\left(\hat{x}_S-\hat{x}_F\right)^2,
\label{eq:HarmonicParticlesHamiltonian}
\end{eqnarray}
with  continuous variables $ \hat{x}_S$ and $\hat{x}_F$ in one spatial dimension. Alternatively, it can be rewritten as
\begin{eqnarray}
\hat{H}_0\left(t\right) & = & \frac{\hat{p}^2_S}{2 m_S} + \frac{\hat{p}^2_F}{2 m_F} + \frac{1}{2} \kappa_S\left(t\right)\hat{x}_S^2 + \frac{1}{2} \kappa_F\left(t\right)\hat{x}_F^2 
\nonumber \\&  & - k_I\left(t\right)\hat{x}_S \hat{x}_F,\label{eq:HarmonicHamiltonian}
\end{eqnarray}
with $\kappa_S= k_S + k_I$ and $\kappa_F= k_F + k_I$, which makes explicit the bilinear coupling. For the sake of generality, we first provide a derivation of the counterdiabatic drivings for  this system when  all spring constants are time-dependent.

The spectral properties  can be studied by diagonalizing the system in terms of two independent harmonic oscillators, the normal modes.
We denote by  $\hat{y}_i$, $\hat{p}_i$, and $\kappa_i$ ($i=1,2$)   the corresponding normal-mode coordinates, conjugate momentum and spring constants, for which explicit expressions are derived in Appendix \ref{sec:TCHO}.
In terms of them, the system Hamiltonian can be simply written as
\begin{eqnarray}
\hat{H}_0\left(t\right) & = & \frac{\hat{p}^2_1}{2 \mu} + \frac{\hat{p}^2_2}{2 \mu} + \frac{1}{2} \kappa_1\left(t\right)\hat{y}_1^2 + \frac{1}{2} \kappa_2\left(t\right)\hat{y}_2^2. \label{eq:NormalModesHamiltonian}
\end{eqnarray}
The BOA leads to an approximation of the system Hamiltonian $\hat{H}_0$ also in terms of two independent harmonic oscillators, whose eigenstates under the conventional and relaxed BOA coincide, as the Berry connection identically vanishes and the quantum geometric tensor reduces to a time-dependent constant; see  Appendix \ref{sec:TCHO} for further details. Note, that in order for the slow sub-system of the BOA Hamiltonian to have a real harmonic frequency it is required that $\kappa_S\left(t\right)\kappa_F\left(t\right) > k_I\left(t\right)^2$.

Knowledge of the exact and BOA eigenfunctions allows us to establish the validity of the BOA whenever  $m_F/m_S\ll1$.
To this end, we consider the fidelity between an exact eigenstate of the two coupled harmonic oscillators $\Psi_{\rm exact}$ and the corresponding BOA $\Psi_{\rm BOA}$ 
\begin{align}
F = \lvert \braket{\Psi_{\rm Exact}} \ket{\Psi_{\rm BOA}} \rvert^2.
\label{eq:Fidelity}
\end{align}
The Born-Oppenheimer method provides a good approximation for the coupled harmonic oscillator problem for a small mass ratio $m_F/m_S$, see Fig.~\ref{fig:Fidelityk}. As  the trapping frequency of any of the two sub-systems -- slow or fast particles -- is increased, the accuracy of the BOA increases, as shown in Fig.~\ref{fig:Fidelityk}(a)-(b). This is consistent with the fact that the energy scale separation between the two subsystems is increased for a given mass ratio  as the sub-system spring constants are made larger, making the two sub-systems more decoupled. By contrast, when decreasing the interaction strength, see Fig~\ref{fig:Fidelityk}(c), the state $\Psi_{\rm BOA}$ approaches $\Psi_{\rm Exact}$ as quantified by the higher values of the fidelity. Naturally, the coupling between both subsystems increases with the interaction spring constant, leading to a breakdown of the BOA at large values of $k_I$.

We next derive and compare the auxiliary control terms required to enforce adiabaticity in an arbitrary prescheduled time using the exact CD and CBOD.

\subsection{Exact Counterdiabatic Driving}

The exact solution of the coupled system (\ref{eq:HarmonicParticlesHamiltonian}) can be written in terms of the two independent harmonic oscillators,  the  normal modes, described by Hamiltonian~\eqref{eq:NormalModesHamiltonian}. Knowledge of the CD term for  a single harmonic oscillator \cite{Muga2010,Campo2013} readily yields the exact CD term for the coupled system
\begin{align}
\hat{H}_1\left(t\right) = 
- \frac{\dot{\omega}_1 \left(t\right)}{4 \omega_1 \left(t\right)} \{ \hat{y}_1 , \hat{p}_1 \}- \frac{\dot{\omega}_2 \left(t\right)}{4 \omega_2 \left(t\right)} \{ \hat{y}_2 , \hat{p}_2 \},
\end{align}
as the sum of  the generators of the squeezing operator for each normal mode. As such, they are spatially non-local due to the momentum dependence.
Alternative controls can be obtained by means of  the unitary transformation 
\begin{align}
\mathcal{U} = \exp \left( - \frac{i \mu \dot{\omega}_1\left(t\right)}{4 \hbar \omega_1\left(t\right)} \hat{y}_1^2 \right) \exp \left( - \frac{i \mu \dot{\omega}_2\left(t\right)}{4 \hbar \omega_2\left(t\right)} \hat{y}_2^2 \right),
\end{align}
which acts on the position and momentum operators in the Hamiltonian as
\begin{eqnarray}
\label{utranseq}
\hat{y}_{1,2} &\rightarrow & \mathcal{U} \hat{y}_{1,2} \mathcal{U}^\dagger =  \hat{y}_{1,2}, \nonumber\\
\hat{p}_{1,2} &\rightarrow & \mathcal{U} \hat{p}_{1,2} \mathcal{U}^\dagger =  \hat{p}_{1,2} + \frac{1}{2} \mu \frac{\dot{\omega}_{1,2}\left(t\right)}{\omega_{1,2}\left(t\right)} \hat{y}_{1,2}, \nonumber\\
\hat{p}_{1,2}^2 &\rightarrow & \mathcal{U} \hat{p}_{1,2}^2 \mathcal{U}^\dagger =  \hat{p}_{1,2}^2 + \frac{\dot{\omega}_{1,2}\left(t\right) \mu}{2\omega_{1,2}\left(t\right)} \{ \hat{y}_{1,2},\hat{p}_{1,2} \} \nonumber\\ & & + \frac{\mu^2 \dot{\omega}_{1,2}\left(t\right)^2}{4 \omega_{1,2}\left(t\right)^2} \hat{y}_{1,2}^2.
\end{eqnarray}
Given that $\partial_t \mathcal{U}^\dagger \neq 0$  the full driving Hamiltonian $\hat{H}=\hat{H}_0+\hat{H}_1$ is transformed according to
\begin{align}
\hat{H} \rightarrow \hat{H}_T=\, \mathcal{U} \hat{H} \mathcal{U}^\dagger - i \hbar \mathcal{U} \partial_t \mathcal{U}^\dagger.
\end{align}
while the original wave function $\Psi$ is mapped to
\begin{align}
\Psi \rightarrow \Psi_T = \mathcal{U} \Psi.
\end{align}
Making use of (\ref{utranseq}), it is found that the transformed Hamiltonian  $\hat{H}_T$, unitarily equivalent to $\hat{H}$, takes the form 
\begin{align}
\hat{H}_T\left(t\right) = \frac{\hat{p}_1^2}{2 \mu} + \frac{\hat{p}_2^2}{2 \mu} + \frac{1}{2} \mu \omega_{T,1}\left(t\right)^2 \hat{y}_1^2 + \frac{1}{2} \mu \omega_{T,2}\left(t\right)^2 \hat{y}_2^2,
\end{align}
with the corresponding frequencies being
\begin{align}
\omega_{T,\{1,2\}}\left(t\right)^2 = \omega_{1,2}\left(t\right)^2 - \frac{3\dot{\omega}_{1,2}\left(t\right)^2}{4 \omega_{1,2}\left(t\right)^2} + \frac{\ddot{\omega}_{1,2}\left(t\right)}{2\omega_{1,2}\left(t\right)}.
\end{align}
Therefore, the exact CD of coupled harmonic oscillators can be implemented by a modification of the normal mode frequency of the original system Hamiltonian $\hat{H}_0$. However, this modification will require independent control of the slow, fast and interaction spring constants.

\subsection{CBOD}

Within the BOA the Hamiltonians of the slow and fast subsystems are that of two harmonic oscillators, and the corresponding CBOD terms  are given by
\begin{align}
\hat{H}_{F,1}\left(t\right) = &- \frac{\dot{\omega}_F \left(t\right)}{4 \omega_F \left(t\right)} \{ \hat{x}_T , \hat{p}_T \},\\
\hat{H}_{S,1}\left(t\right) = & -\frac{\dot{\omega}_S \left(t\right)}{4 \omega_S \left(t\right)} \{ \hat{x}_S , \hat{p}_S \},
\end{align}
where $\hat{x}_T = \left(\hat{x}_F - \frac{k_I\left(t\right)}{\kappa_F\left(t\right)} x_S\right)$ and $\hat{p}_T=\hat{p}_F$ is the corresponding conjugate momentum operator, see appendix  \ref{sec:TCHO}. For the coupled harmonic oscillators the CBOD auxiliary controls under  the conventional  and relaxed BOA are equivalent, as the wave functions coincide. We will consider a case in which the fast sub-system is also driven to enforce adiabaticity by CBOD. However, within the BOA the fast sub-system is usually assumed to evolve adiabatically and, in that case, the control $\hat{H}_{F,1}$ would not be required. 
The driving Hamiltonian  with the CBOD control terms reads
\begin{eqnarray}
& & \hat{H}\left(t\right) = \frac{\hat{p}^2_S}{2 m_S} + \frac{\hat{p}^2_F}{2 m_F} + \frac{1}{2} \kappa_S\left(t\right)\hat{x}_S^2 + \frac{1}{2} \kappa_F\left(t\right)\hat{x}_F^2 \nonumber 
\\ & & -  k_I\left(t\right)\hat{x}_S\hat{x}_F
- \frac{\dot{\omega}_S}{4 \omega_S} \{ \hat{x}_S,\hat{p}_S \}- \frac{\dot{\omega}_F}{4 \omega_F} \{ \hat{x}_T,\hat{p}_F \}.\label{eq:StartingCBODHamiltonian}
\end{eqnarray}
The evolution under this Hamiltonian is not necessarily adiabatic with respect to the exact $\hat{H}_0$ eigenbasis, as the CBOD auxiliary terms are approximate. As a result, an STA designed by CBOD can not be arbitrarily fast. 
The direct exact solution of the Schr\"{o}dinger equation with this Hamiltonian is hindered by the term $\hat{{x}}_S \hat{{p}}_F$ resulting from the last anti-commutator. However, this term can be absorbed  into the fast momentum. Dropping the time dependence for simplicity, one finds
\begin{eqnarray}
\frac{\hat{p}^2_F}{2 m_F} + \frac{\dot{\omega}_F k_I}{2 \omega_F \kappa_F} \hat{x}_S \hat{p}_F & =  & \frac{1}{2 m_F} \left( \hat{p}_F + m_F \frac{\dot{\omega}_F k_I}{2 \omega_F \kappa_F} \hat{x}_S\right)^2 \nonumber \\ & & - m_F \frac{\dot{\omega}_F^2 k_I^2}{8 \omega_F^2 \kappa_F^2} \hat{x}_S^2 \nonumber \\
& \approx & \frac{\hat{p}_F^2}{2 m_F}- m_F \frac{\dot{\omega}_F^2 k_I^2}{8 \omega_F^2 \kappa_F^2} \hat{x}_S^2,\label{eq:MomentumApproximation}
\end{eqnarray}
where in the final line we have made an approximation consistent with the BOA, that the fast sub-system momentum will dominate over the additional momentum term which is a function of the slow sub-system coordinate. The driving Hamiltonian takes then the form
\begin{eqnarray}
\hat{H}\left(t\right) & = & \frac{\hat{p}^2_S}{2 m_S} + \frac{\hat{p}^2_F}{2 m_F} + \frac{1}{2} \left(\kappa_S\left(t\right)- m_F \frac{\dot{\omega}_F^2 k_I^2}{4 \omega_F^2 \kappa_F^2} \right)\hat{x}_S^2 \nonumber\\&  & + \frac{1}{2} \kappa_F\left(t\right)\hat{x}_F^2 -  k_I\left(t\right)\hat{x}_S\hat{x}_F
- \frac{\dot{\omega}_S}{4 \omega_S} \{ \hat{x}_S,\hat{p}_S \} \nonumber\\&  & - \frac{\dot{\omega}_F}{4 \omega_F} \{ \hat{x}_F,\hat{p}_F \}.
\label{eq:ApproxCBODHamiltonian}
\end{eqnarray}

In a similar manner to the previous scenario, we can use the unitary transformation of 
\begin{align}
\mathcal{U} = & \exp\left[ - i \left(\frac{ m_S \dot{\omega}_S\left(t\right)}{4 \hbar \omega_S\left(t\right)} x_S^2  + \frac{ m_F \dot{\omega}_F\left(t\right)}{4 \hbar \omega_F\left(t\right)} x_F^2 \right) \right]
\end{align}
to obtain the unitarily equivalent Hamiltonian
\begin{align}
\hat{H}\left(t\right) = & \frac{\hat{p}^2_S}{2 m_S} + \frac{\hat{p}^2_F}{2 m_F} + \frac{1}{2} \gamma_S\left(t\right) \hat{x}_S^2 + \frac{1}{2} \gamma_F\left(t\right)\hat{x}_F^2 \nonumber \\ & -  k_I\left(t\right)\hat{x}_S\hat{x}_F \label{eq:CBODHamiltonian}
\end{align}
which is that of two oscillators with bilinear coupling, i.e., the original system Hamiltonian \eqref{eq:ApproxCBODHamiltonian},  with modified spring constants
\begin{eqnarray}
\gamma_S &= & \kappa_S - \frac{3 m_S \dot{\omega}_S^2}{4 \omega_S^2} - \frac{m_F \dot{\omega}_F^2 k_I^2}{4 \omega_F^2 \kappa_F^2} + \frac{m_S \ddot{\omega}_S}{2\omega_S}, \\
\gamma_F & = & \kappa_F - \frac{3 m_F \dot{\omega}_F^2}{4 \omega_F^2} + \frac{m_F \ddot{\omega}_F}{2\omega_F}.
\end{eqnarray}
Therefore,  CBOD simplifies the engineering of STA in the system by driving the slow and fast sub-systems independently. In addition,  it succeeds in doing so  without the need to  tailor the interaction term between the two-subsystems. This is contrary to the exact counterdiabatic driving, which involves a controlled modulation in time of all the potential terms in the original Hamiltonian, including the interaction. 

\begin{figure*}[t]
\begin{center}        
        \includegraphics[width=0.98\linewidth]{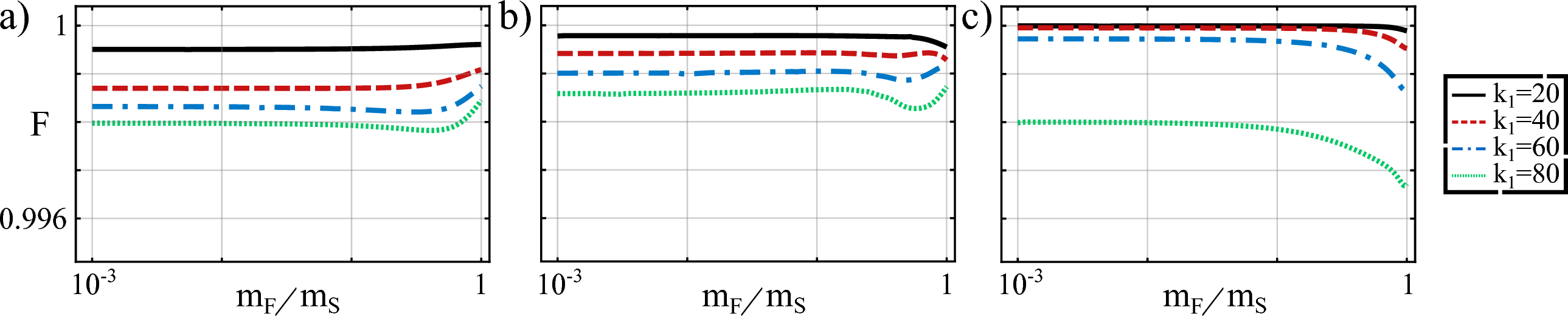}
\end{center}
\caption{\label{fig:FidelityBOA}
{\bf Fidelity under Counterdiabatic Born-Oppenheimer Dynamics (CBOD).} The ramping of a single spring constant of two coupled harmonic oscillators is considered, according to the  time modulation  in Eq. (\ref{eq:Ramp}).  Under exact CD, a system initialized in the ground state  evolves into  the ground-state of the final Hamiltonian upon completion of the ramp. CBOD approximates the required controls to assist adiabaticity, facilitating their implementation.
The fidelity between the final state evolved under the exact CD and CBOD is shown (for $t=T_f=1$) in three different cases, corresponding to the modulation of one of each of the spring constants in the systems Hamiltonian: a) Ramping $\kappa_S$, with $k_I=50,\kappa_F=100,k_0=50$. b) Ramping $\kappa_F$, with $k_I=50,\kappa_S=100,k_0=50$. c) Ramping $k_I$, with $\kappa_S=\kappa_F=100,k_0=1$.}
\end{figure*}

\begin{figure*}[t]
\begin{center}        
        \includegraphics[width=0.98\linewidth]{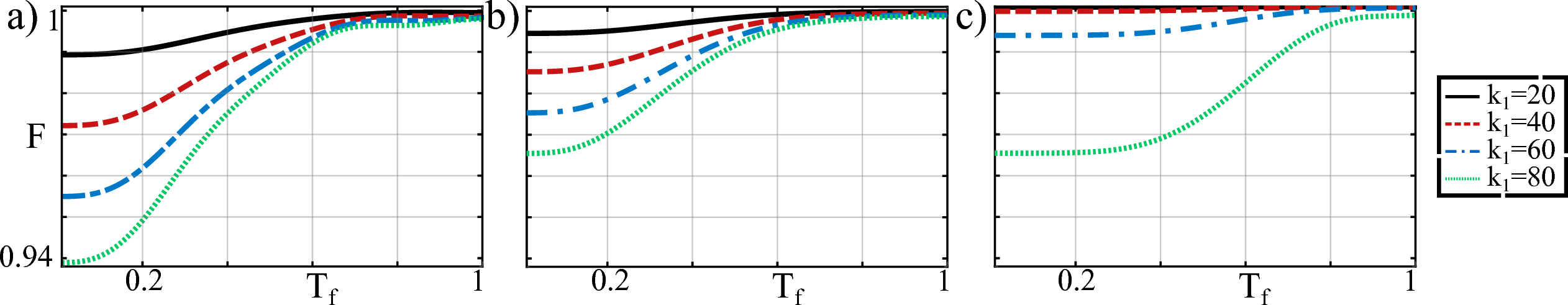}
\end{center}
\caption{\label{fig:FidelityTime}
{\bf Speed of Counterdiabatic Born-Oppenheimer Dynamics (CBOD).} The ramping of a single spring constant of two coupled harmonic oscillators is considered as a function of the ramping time $T_f$, according to the  time modulation  in Eq. (\ref{eq:Ramp}).  
The fidelity between the final state evolved under the exact CD and CBOD is shown in three different cases (for $t=T_f$), corresponding to the modulation of one of each of the spring constants in the systems Hamiltonian: a) Ramping $\kappa_S$, with $k_I=50,\kappa_F=100,k_0=50$. b) Ramping $\kappa_F$, with $k_I=50,\kappa_S=100,k_0=50$. c) Ramping $k_I$, with $\kappa_S=\kappa_F=100,k_0=1$. }
\end{figure*}

As CBOD relies on the BOA, the dynamics generated by the Hamiltonian \eqref{eq:CBODHamiltonian} is not strictly adiabatic. The exact solution to the corresponding time-independent Schr\"{o}dinger equation can be obtained in a similar manner to that shown for the exact solution in Appendix \ref{sec:TCHO}, with a separation into two independent normal-mode harmonic oscillators. The non-adiabatic  evolution of each normal-mode harmonic oscillator can then be described exactly by a self-similar transformation of the corresponding wave function  $\Phi$ at the start of the evolution. For a harmonic oscillator of mass $m$ and frequency $\omega\left(t\right)$ the scaling symmetry determines the evolution of the ground state according to \cite{Muga2010,Chen2010a,Campo2013,Arimondo2013Chap}
\begin{align}
\Phi\left(t\right) = & \frac{1}{\sqrt{b\left(t\right)}} \exp\left( i \frac{m \dot{b}\left(t\right)}{2 \hbar b\left(t\right)} x^2 - i \frac{\omega_0}{2} \int_0^t dt^\prime \frac{1}{b\left(t^\prime\right)^2}  \right) \times \nonumber \\ & \Phi\left(\frac{x}{b\left(t\right)};0\right),\label{eq:HORescaling}
\end{align}
where $\omega_0 = \omega\left(0\right)$ and $b\left(t\right)>0$ is a scaling factor obtained by solving the Ermakov equation
\begin{align}
\ddot{b}\left(t\right) + \omega\left(t\right)^2 b\left(t\right)^2 = \frac{\omega_0^2}{b\left(t\right)^3},
\end{align}
with boundary conditions $b(0)=1$ and $\dot{b}(0)=0$. By solving the above Ermakov equation numerically for the parameters of Hamiltonian~\eqref{eq:CBODHamiltonian},  the exact evolution of the system under CBOD can be obtained via Eq.~\eqref{eq:HORescaling}.

\subsection{CBOD Validity}

To investigate the validity of the CBOD technique we consider the following modulation of the spring constant 
\begin{align}
K\left(t\right) = k_0 + \frac{k_1}{T_f} \left[t - \frac{T_f}{2 \pi} \sin\left(\frac{2\pi}{T_f} t\right) \right],
\label{eq:Ramp}
\end{align}
where $k_0$ and $k_1$ are offset and strength parameters of the ramping, respectively, and $T_f$ is the time-scale of the modulation, with a single modulation after $t=T_f$. This ramp has first and second derivatives that vanish at $t=0$ and $T_f$, favoring adiabatic evolution \cite{Masuda2014}, e.g., over a  linear ramp. We will consider a ramping of the individual terms in Hamiltonian~\eqref{eq:HarmonicParticlesHamiltonian}, i.e. $\kappa_S$, $\kappa_F$ and $k_I$. 

We show the fidelity of the ground state under a ramping of each spring constant in Fig.~\ref{fig:FidelityBOA}, i.e.
\begin{align}
F = \lvert \braket{\Psi_{\rm Exact}\left(T_f\right)} \ket{\Psi_{\rm CBOD}\left(T_f\right)} \rvert^2.
\end{align}
From the assessment of the validity of the BOA in this example, see Fig.~\ref{fig:Fidelityk}, the CBOD technique is expected to be particularly sensitive to ramping the interaction spring constant. A significant drop off in fidelity  is observed  as the interaction strength  is increased by the ramp. Overall, CBOD matches with high fidelity  the exact adiabatic evolution. This is reflected by the values $F\geq0.99$ observed in Fig.~\ref{fig:FidelityBOA}.

For large ramps of the interaction spring constant, the dotted green lines of Fig.~\ref{fig:FidelityBOA}, a slight breakdown of the validity of CBOD for $m_F/m_S \sim 1$ becomes manifest. This is most likely due to the momentum approximation of Eq.~\eqref{eq:MomentumApproximation}, which was required to decouple the fast momentum and slow coordinates in the CD control terms within the BOA. This approximation is only valid in the limit of $m_F/m_S \ll 1$ and/or $k_I \ll \kappa_F$, which are both broken in this scenario. Hence, the lower fidelity of the CBOD protocol for large mass ratio. The breakdown of the CBOD for $m_F/m_S \sim 1$ is consistent with the breakdown of the BOA. Provided that the fast and slow degrees of freedom can so be defined,  CBOD can generate high fidelity evolution of states under fast modulations.

However, by contrast to exact CD,  CBOD is not  valid for arbitrarily fast modulations.  CBOD trades the possibility of  engineering arbitrarily fast STA for the ability to treat interacting systems which may not be exactly solvable and to simplify the experimental implementation of the required control terms in these systems. The behavior of CBOD under faster modulations is investigated in Fig.~\ref{fig:FidelityTime}, by the fidelity of the exact and CBOD techniques at the end of the process $t=T_f$. As would be expected, the fidelity decreases with faster ramping times, $T_f$ but the fidelity remains high ($F\geq0.9$) and this trend continues for smaller $T_f$, for which the fidelity exhibits  a plateau. In Fig.~\ref{fig:FidelityTime}, each ramp is of a moderate strength and as this strength is increased the fidelity decreases, as was shown in Fig.~\ref{fig:FidelityBOA}. Therefore, despite the dependence of the fidelity  on the modulation time, CBOD can be used for fast driving of the system.

\section{CBOD in trapped charged particles}\label{CBODexCharge_sec}
CBOD can be used to design of STA in systems that are not exactly solvable, or more generally, easily tractable, analytically or numerically.
To demonstrate this,  we consider two particles interacting via a Coulomb potential, with the slow particle  being confined by a harmonic trap and the fast particle feeling only an attractive Coulomb-like interaction. This model is similar to that of Hooke's atom, which is exactly solvable for certain parameter values. The model Hamiltonian is
\begin{align}
\hat{H} = \frac{\hat{\mathbf{p}}_S^2}{2 m_S} + \frac{\hat{\mathbf{p}}_F^2}{2 m_F} + \frac{1}{2} m_S \omega_S^2 \hat{r}_S^2 - \frac{g}{|\hat{r}_F - \hat{r}_S|},
\end{align}
where $g$ is an interaction strength and $\hat{r}$ is the radial coordinate of the spherical coordinate system for each of the slow and fast particles. To justify a separation of variables under BOA, we assume that $m_S \gg m_F$. In this section, we will derive the form of the CBOD drivings for this model.

Under the BOA, the slow and fast Hamiltonian are
\begin{align}
\hat{H}_S = & \frac{\hat{\mathbf{p}}_S^2}{2 m_S} + \frac{1}{2} m_S \omega_S^2 \hat{r}_S^2 + \varepsilon\left(\hat{r}_S\right), \\
\hat{H}_F = & \frac{\hat{\mathbf{p}}_F^2}{2 m_F} - \frac{g}{|\hat{r}_F - \hat{r}_S|}.
\end{align}
The fast sub-system has the form of the hydrogen atom Hamiltonian which has known solutions, 
\begin{align}
\ket{\psi_F\left(r_F^\prime,\theta,\phi\right)} = \ket{R_{n,l}\left(r_F^\prime\right)} \otimes \ket{Y_{l,m}\left(\theta,\phi\right)},
\end{align}
where $\hat{r}_F^\prime = \hat{r}_F - \hat{r}_S$.
The solutions to the radial $\ket{R_{n,l}\left(r_F^\prime\right)}$ and angular $\ket{Y_{l,m}\left(\theta,\phi\right)}$ separation of this wavefunction can be found by solving the separated Schr\"{o}dinger equations, see Refs.~\cite{Coulson1958,Brandt2012,weinberg2015Book}, and are characterized by three quantum numbers $(n,l,m)$. The corresponding eigenvalues depend only on the principal quantum number $n$ and take the form
\begin{align}
\varepsilon = - \frac{m_F g^2}{2 \hbar^2 n^2}.
\end{align}
The reduced Hamiltonian thus becomes
\begin{align}
\hat{H}_S =  \frac{\hat{\mathbf{p}}_S^2}{2 m_S} + \frac{1}{2} m_S \omega_S^2 \hat{r}_S^2 - \frac{m_F g^2}{2 \hbar^2 n^2},
\end{align}
which has harmonic oscillator solutions with energy
\begin{align}
E_{u,n} = \hbar \omega_S \left( u +\frac{3}{2} \right) - \frac{m_F g^2}{2 \hbar^2 n^2}.
\end{align}

We consider the driving of the  system by modulating the interaction strength $g = g(t)$. Under the BOA the dynamics arises only in the fast sub-system, as the slow sub-systems state is invariant under a driving of $g$. The $g$-dependence of the fast sub-system state is entirely contained within the radial component of the wavefunction \cite{Coulson1958,Woan2000,Brandt2012}, which takes the normalised form
\begin{align}
\ket{R_{n,l}\left(r_F^\prime\right)} = & \left(\frac{2 m_F g}{\hbar^2 n}\right)^\frac{3}{2} \sqrt{\frac{(n-l-1)!}{2n(n+l)!}} \exp \left(-\frac{m_F g r_F^\prime}{\hbar^2 n}\right) \nonumber \\ & \times \left(\frac{2 m_F g r_F^\prime}{\hbar^2 n}\right)^l L^{2l+1}_{n-l-
1}\left(\frac{2 m_F g r_F^\prime}{\hbar^2 n} \right) ,
\end{align}
where $L_m^\alpha(x)$ are the generalised Laguerre polynomials. In this scenario, it is helpful to write the CD term as
\begin{align}
\hat{H}_{1,F} \left(t\right) = i \hbar \sum_{n,l} \big( & \dot{g} \ket{\partial_g R_{n,l}} \bra{R_{n,l}} \nonumber \\ & - \dot{g} \braket{R_{n,l}}\ket{\partial_g R_{n,l}} \ket{R_{n,l}} \bra{R_{n,l}} \big) .
\label{eq:HydrogenCD}
\end{align}
We obtain the $g$ derivative of the radial component as
\begin{align}
\dot{g}\ket{\partial_g R_{n,l}\left(r_F^\prime\right)} & = \bigg[ \frac{3}{2} - \frac{m_F g r_F^\prime}{\hbar^2 n} + (n-1) \nonumber \\ & - (n+l) \frac{L_{n-l-2}^{2l +1}\left(\frac{2 m_F g r_F^\prime}{\hbar^2 n}\right)}{L_{n-l-1}^{2l+1}\left(\frac{2 m_F g r_F^\prime}{\hbar^2 n}\right)} \bigg] \frac{\dot{g}}{g}\ket{R_{n,l}\left(r_F^\prime\right)},
\label{eq:RadialD}
\end{align}
where we have simplified the expression using the generalised Laguerre polynomial recurrence relations \cite{Gradshteyn2014}. The Berry connection, $\dot{g}\braket{R_{n,l}}\ket{\partial_g R_{n,l}}$, takes the form of a known definite integral \cite{Prudnikov1986} and is found to be
\begin{align}
\dot{g}\braket{R_{n,l}}& \ket{\partial_g R_{n,l}} = \frac{\dot{g}}{g} \bigg[  \frac{1}{2} - \frac{n}{2} - \frac{l\left(l+1\right)}{2n}  \nonumber \\ & - \frac{2 g^2 m_F^2 (n+1)}{\hbar^4 n^3} \frac{\Gamma\left(2l+2\right) \left(1\right)_{n-l-2} \left(2l+2\right)_{n-l-1}}{\left(n-l-2\right)! \left(n+l\right)!}\bigg],
\label{eq:HydrogenBerry}
\end{align}
with $\Gamma(a)$ the gamma function and $\left(a\right)_{k}$ the Pochhammer symbol, i.e. $\left(a\right)_{k} = \Gamma(a+k)/\Gamma(a)$. Using Eqs.~\eqref{eq:RadialD} and~\eqref{eq:HydrogenBerry} it is possible to construct the CBOD fast counter-diabatic drivings in general. For the sake of illustration, we consider the system to be prepared in the low energy states. We can compactly write the form for the CD of single states for the ground state $(n,l) = (1,0)$
\begin{align}
H_{F,1} = i \hbar \left( \frac{3}{2} - \frac{m_F g r_F^\prime}{\hbar^2 }\right) \frac{\dot{g}}{g}\ket{R_{1,0}} \bra{R_{1,0}},
\end{align}
which is an energy term plus a potential linear in the radial coordinate. 
The first excited state is degenerate with $(n,l)=(2,0)$ or $(n,l)=(2,1)$, and each has a different CD term, for $(2,0)$
\begin{align}
H_{F,1} = & i \hbar \bigg( 3 - \frac{g m_F \left( g m_F + \hbar^2 r_F^\prime \right)}{2 \hbar^4} \nonumber \\ & - \frac{2 \hbar^2}{\hbar^2 - g m_F r_F^\prime}\bigg) \frac{\dot{g}}{g}\ket{R_{2,0}} \bra{R_{2,0}},
\end{align}
and for $(2,1)$
\begin{align}
H_{F,1} = i \hbar \left( \frac{7}{2} - \frac{m_F g r_F^\prime}{2\hbar^2 }\right) \frac{\dot{g}}{g}\ket{R_{2,1}} \bra{R_{2,1}}.
\end{align}
We see that for higher energy states, the form of the CD, Eq.~\eqref{eq:HydrogenCD}, required can be more complex, with $1/r_F^\prime$ potentials for the $(2,0)$ state, as the CD has a term proportional to the ratio of two Laguerre polynomials from the derivative Eq.~\eqref{eq:RadialD}. 

This example illustrates how CBOD can prove useful to engineer STA in complex systems. In this particular model, the full Hamiltonian is not easily solvable. Yet, the derivation of the CBOD auxiliary controls is made possible by relating the subsystem Hamiltonians to well-known solvable models. More generally,  we expect that CBOD helps cracking the complexity barrier in the design of STA   by harnessing the separation of energy scales between different degrees of freedom, whenever present.


\section{Conclusions}

Shortcuts to Adiabaticity (STA)  provide control protocols to guide the dynamics of quantum and classical systems along an adiabatic reference trajectory, without relying on slow driving.  A universal approach to designing STA is provided by the counterdiabatic driving  (CD) technique that guides the evolution of an arbitrary quantum system by means of auxiliary control fields. However, determining the auxiliary controls requires knowledge of the spectral properties of the system, hindering its application to complex systems. 

In this work, we have introduced Counterdiabatic Born-Oppenheimer Dynamics (CBOD) as a framework to design STA in  complex systems. CBOD identifies the required controls to speed up the dynamics of the system by invoking the Born-Oppenheimer approximation (BOA) whenever a separation between fast and slow degrees of freedom is justified. In such a scenario, the required CD terms for the fast and slow variables can be obtained in two subsequent steps, which in the spirit of the BOA, avoids the need to diagonalize the high-dimensional Hamiltonian of the full system.  Thus, CBOD  facilitates the finding of the required Hamiltonian  controls to speed up the dynamics, in scenarios where spectral properties are not readily available. In addition, CBOD also simplifies the implementation of the STA by reducing the need to control the coupling between fast and slow degrees of freedom. We have demonstrated the validity of CBOD by testing it in a paradigmatic test-bed of BOA, an exactly-solvable model of two driven coupled harmonic oscillators with unequal masses for which CBOD competes with the exact counterdiabatic driving in the preparation of a target state. We have also applied CBOD to the design STA in a more complex Coulomb system.
We anticipate that the CBOD technique should facilitate the fast nonadiabatic control of the dynamics of  complex systems  in the plethora of scenarios in which the Born-Oppenheimer approximation has proved useful.


\section*{Acknowledgements}

It is a pleasure to thank Luis Pedro Garc\'{i}a-Pintos and Stuart A. Rice for insightful discussions and comments on the manuscript. C.W.D. acknowledges studentship funding from EPSRC CM-CDT Grant No. EP/L015110/1 and support from SUPA under the Postdoctoral and Early Career Researcher Exchange Program. C.W.D. thanks the University of Massachusetts Boston for their hospitality during this work. Funding from the John Templeton Foundation and UMass Boston (project P20150000029279) is further acknowledged.

\appendix

\section{Two coupled harmonic oscillators with unequal masses}
\label{sec:TCHO}
\subsection{Exact solution}
\label{sec:HOExact}

The exact solution to the Schr\"{o}dinger equation of Hamiltonian~\eqref{eq:HarmonicHamiltonian} is well known \cite{Han1999}. First, we transform the position and momentum spaces canonically via the transformations
\begin{align}
\begin{pmatrix}
\hat{p}_1 \\
\hat{p}_2
\end{pmatrix} = \begin{pmatrix}
\left(m_F/m_S\right)^{\frac{1}{4}} & 0 \\
0 & \left(m_S/m_F\right)^{\frac{1}{4}}
\end{pmatrix} \begin{pmatrix}
\hat{p}_S \\
\hat{p}_F
\end{pmatrix},
\end{align}
and
\begin{align}
\begin{pmatrix}
\hat{x}_1 \\
\hat{x}_2
\end{pmatrix} = \begin{pmatrix}
\left(m_S/m_F\right)^{\frac{1}{4}} & 0 \\
0 & \left(m_F/m_S\right)^{\frac{1}{4}}
\end{pmatrix} \begin{pmatrix}
\hat{x}_S \\
\hat{x}_F
\end{pmatrix}.
\end{align}
This is followed by a rotation of the coordinates 
\begin{align}
\begin{pmatrix}
\hat{y}_1 \\
\hat{y}_2
\end{pmatrix} = \begin{pmatrix}
\cos\alpha & -\sin\alpha \\
\sin\alpha & \cos\alpha
\end{pmatrix} \begin{pmatrix}
\hat{x}_1 \\
\hat{x}_2
\end{pmatrix},
\end{align}
under which the momentum is invariant. To diagonalize the system the rotation angle is found to be
\begin{align}
\alpha\left(t\right) = \frac{1}{2} \arctan \left( \frac{2 k_I\left(t\right)}{\kappa_S\left(t\right) \sqrt{\frac{m_F}{m_S}} - \kappa_F\left(t\right) \sqrt{\frac{m_S}{m_F}}} \right).
\end{align}
The Hamiltonian is then diagonalized into normal modes, i.e.,  two independent harmonic oscillators 
\begin{eqnarray}
\hat{H}_0\left(t\right) & = & \frac{\hat{p}^2_1}{2 \mu} + \frac{\hat{p}^2_2}{2 \mu} + \frac{1}{2} \kappa_1\left(t\right)\hat{y}_1^2 + \frac{1}{2} \kappa_2\left(t\right)\hat{y}_2^2, \label{eq:NormalModesHamiltonianApp}
\end{eqnarray}
with reduced mass $\mu = \sqrt{m_S m_F}$ and spring constants
\begin{eqnarray}
\kappa_1\left(t\right)& = &\sqrt{\frac{m_F}{m_S}} \kappa_S\left(t\right) \cos^2\alpha + \sqrt{\frac{m_S}{m_F}} \kappa_F\left(t\right) \sin^2\alpha \nonumber\\ & &+ 2 k_I\left(t\right) \sin\alpha\cos\alpha, \\
\kappa_2\left(t\right)& = &\sqrt{\frac{m_F}{m_S}} \kappa_S\left(t\right) \sin^2\alpha + \sqrt{\frac{m_S}{m_F}} \kappa_F\left(t\right) \cos^2\alpha \nonumber\\ & &- 2 k_I\left(t\right) \sin\alpha\cos\alpha.
\end{eqnarray}
The solution to the time-independent Schr\"{o}dinger equation of Hamiltonian~\eqref{eq:NormalModesHamiltonianApp} is that of two independent harmonic oscillators in the coordinates ${y}_1$ and ${y}_2$ with total energy
\begin{align}
\epsilon_{i j}\left(t\right) = \hbar \omega_1\left(t\right) \left(i + \frac{1}{2}\right) + \hbar \omega_2\left(t\right) \left(j + \frac{1}{2}\right),
\end{align}
with $i,j=0,1,2,\dots$ and frequencies $\omega_{1,2}\left(t\right) = \sqrt{\kappa_{1,2}\left(t\right)/\mu}$. The full wave functions take the form of the usual Harmonic oscillator solutions, i.e.
\begin{widetext}
\begin{align}
\psi_{i j} \left(y_1,y_2\right) = \frac{1}{\sqrt{2^{i+j} i! j!}} \left(\frac{\mu^2 \omega_1 \omega_2}{\left(\pi \hbar\right)^2}\right)^{\frac{1}{4}} \exp\left(- \frac{\mu}{2 \hbar} \left( \omega_1 y_1^2 + \omega_2 y_2^2 \right) \right) H_i\left(\sqrt{\frac{\mu \omega_1}{\hbar}} y_1\right) H_j\left(\sqrt{\frac{\mu \omega_2}{\hbar}} y_2\right).
\end{align}
\end{widetext}
With the exact driving utilised in the main text the time evolution will be the adiabatic solution
\begin{align}
\psi_{i j} \left(y_1,y_2,t\right) = \exp\left(-\frac{i}{\hbar} \int_0^t dt^\prime \epsilon_{i j}\left(t^\prime\right) \right)\psi_{i j} \left(y_1,y_2\right).
\end{align}

\subsection{Born-Oppenheimer approximation for two coupled harmonic oscillators}

We now consider Hamiltonian~\eqref{eq:HarmonicHamiltonian} in the regime of $m_S \gg m_F$, where the BOA is valid. Following the steps of Sec.~\ref{sec:BOA}, the time-independent Schr\"{o}dinger equation of the fast subsystem reads
\begin{align}
\left[ \frac{\hat{p}_T^2}{2 m_F} + \frac{1}{2} \kappa^{\prime}_S\left(t\right) x_S^2 + \frac{1}{2} \kappa_F\left(t\right) \hat{x}_T^2\right] &\phi_n\left(x_F;x_S\right) \nonumber \\  = \varepsilon_n\left(x_S\right) \phi_n\left(x_F;x_S\right)&,\label{eq:HarmonicBOAFastHamiltonian}
\end{align}
with transformed coordinate $\hat{x}_T = \left(\hat{x}_F - \frac{k_I\left(t\right)}{\kappa_F\left(t\right)} x_S\right)$, momentum $\hat{p}_T = \hat{p}_F$ and slow sub-system spring constant $\kappa^{\prime}_S\left(t\right) = \kappa_S\left(t\right) - \frac{k_I\left(t\right)^2}{\kappa_F\left(t\right)}$. Eq. ~\eqref{eq:HarmonicBOAFastHamiltonian} has solutions of a harmonic oscillator in the $x_T$ coordinate,
\begin{align}
\phi_{n} \left(x_T\right) = & \frac{1}{\sqrt{2^{n} n!}} \left(\frac{m_F \omega_F}{\pi \hbar}\right)^{\frac{1}{4}} \exp\left(- \frac{m_F}{2 \hbar} \omega_F x_T^2 \right) \nonumber \\ & \times H_n\left(\sqrt{\frac{m_F \omega_F}{\hbar}} x_T\right),
\label{eq:FastSol}
\end{align}
with eigenvalues
\begin{align}
\varepsilon_n\left(x_S\right) = \hbar \omega_F\left(t\right) \left(n+\frac{1}{2}\right) + \frac{1}{2} \kappa^{\prime}_S\left(t\right) x_S^2,
\end{align}
and frequency $\omega_F\left(t\right) = \sqrt{\kappa_F\left(t\right)/m_F}$.

The slow sub-system then follows the Schr\"{o}dinger equation of 
\begin{align}
\left[\frac{\hat{p}_S^2}{2 m_S} + \hbar \omega_F\left(t\right) \left(n+\frac{1}{2}\right) + \frac{1}{2} \kappa^{\prime}_S\left(t\right) \hat{x}_S^2\right]  \psi\left({x}_S\right) & \nonumber \\ = E \psi\left({x}_S\right), &
\end{align}
which will have harmonic solutions in ${x_S}$
\begin{align}
\psi_{v} \left(x_S\right) = & \frac{1}{\sqrt{2^{v} v!}} \left(\frac{m_S \omega_S}{\pi \hbar}\right)^{\frac{1}{4}} \exp\left(- \frac{m_S}{2 \hbar} \omega_S x_S^2 \right) \nonumber \\ & \times H_v\left(\sqrt{\frac{m_S \omega_S}{\hbar}} x_S\right).
\end{align}
with frequency $\omega_S = \sqrt{\kappa_S^\prime\left(t\right)/m_S}$. Note, that $\kappa_S^\prime\left(t\right)$ can be negative, turning the frequency imaginary and the solutions considered in this work incorrect. We will take care to ensure that we remain in the real frequency limit, i.e. $\kappa_S\left(t\right)\kappa_F\left(t\right) > k_I\left(t\right)^2$. The total energy of the system will be
\begin{align}
E_{n,v} =  \hbar \omega_S\left(t\right) \left(v+\frac{1}{2}\right)+ \hbar \omega_F\left(t\right) \left(n+\frac{1}{2}\right),
\end{align}
and total wave function for a single modes is
\begin{align}
\Psi_{u v} = \phi_u \psi_v.
\end{align}

\subsection{Relaxed Born-Oppenheimer approximation}

We next consider the application of the  relaxed BOA to  the coupled harmonic oscillators. The Berry connection in Eq.~\eqref{eq:GeneralBOVector}  and geometric tensor given by Eq.~\eqref{eq:GeneralBOScalar} are both integrals of the fast (or reduced) sub-system wave functions, $\phi_u$, which are given in Eq.~\eqref{eq:FastSol}. In particular, we consider the fast sub-system to be in the ground state,
\begin{align} 
\phi_0 = \left(\frac{m_F \omega_F}{\pi \hbar}\right)^{\frac{1}{4}} \exp\left(-\frac{m_F \omega_F}{2\hbar}
x_T^2\right).
\end{align}
The Berry connection identically vanishes
\begin{align}
\mathcal{A}_0 = i \braket{\phi_0}\ket{\partial_{x_S} \phi_0} = 0,
\end{align}
and the geometric tensor simply reads
\begin{align}
g_0 = \langle \partial_{x_S}\phi_0|\partial_{x_S}\phi_0\rangle= \frac{k_I\left(t\right)^2 \omega_F\left(t\right) m_F}{2 \hbar \kappa_F\left(t\right)^2}.
\end{align}
Excited states of the fast sub-system also result in a vanishing Berry connection and a geometric tensor of similar form to above. Therefore within the relaxed BOA, the coupled oscillator problem has the same wave functions as the conventional BOA but with a different total energy. That difference is the value of the geometric tensor multiplied by a factor, as shown in Eq.~\eqref{eq:BOSchrodingerVector}. In this example, the relaxed and conventional counterdiabatic drivings are identical, as the wave functions coincide.


%

\end{document}